\RequirePackage{lineno}
\documentclass[aps, prl,
floatfix,twocolumn,superscriptaddress,
groupedaddress,
showpacs]{revtex4-1}
\usepackage{t1enc}
\usepackage[english]{babel}
\usepackage{graphicx}
\usepackage{epsfig}
\usepackage{amssymb}
\usepackage{dcolumn}% Align table columns on decimal point
\usepackage{bm}% bold math
\usepackage{color}
\usepackage{float}
\usepackage{natbib}
\usepackage{footnote}

\newcommand{\BFCA}{\ensuremath{\mathrm{Ba(Fe}_{1-x}\mathrm{Co}_{x})_{2}\mathrm{As}_{2}}}

\newcommand{\BFAP}{\ensuremath{\mathrm{BaFe}_{2}{\mathrm{(As}_{1-x}\mathrm{P}_{x}}\mathrm{)}_2}}
\newcommand{\EFAP}{\ensuremath{\mathrm{EuFe}_{2}{\mathrm{(As}_{1-x}\mathrm{P}_{x}}\mathrm{)}_2}}

\newcommand{\LFA}{\ensuremath{\mathrm{Li}\mathrm{Fe}\mathrm{As}}}
\newcommand{\NFCA}{\ensuremath{\mathrm{NaFe}_{1-x}\mathrm{Co}_{x}\mathrm{As}}}
\newcommand{\NFRA}{\ensuremath{\mathrm{NaFe}_{1-x}\mathrm{Rh}_{x}\mathrm{As}}}

\newcommand{\ybco}{\ensuremath{\mathrm{YBa}_2\mathrm{Cu}_3\mathrm{O}_{7-\delta}}}

\begin{document}

\title{Non-Fermi-liquid scattering rates and anomalous band dispersion in ferropnictides}
%\author{J. Fink}
%\affiliation{Leibniz Institute for Solid State and Materials Research  Dresden, Helmholtzstr. 20, D-01069 Dresden, Germany}
%\author{A. Charnukha}
%\affiliation{Leibniz Institute for Solid State and Materials Research  Dresden,\,Helmholtzstr. 20,~D-01069 Dresden,~Germany}
%\author{E.D.L. Rienks}
%\affiliation{\nobreak{Helmholtz-Zentrum Berlin,~Albert-Einstein-Strasse 15,~D-12489 Berlin,~Germany}}
%\author{S. Thirupathaiah}
%\affiliation{Helmholtz-Zentrum Berlin, Albert-Einstein-Strasse 15, D-12489 Berlin, Germany}
%\author{I. Avigo}
%\affiliation{Fakult\"at f\"ur Physik, Universit\"at Duisburg-Essen, Lotharstr. 1, D-47075 Duisburg, Germany}
\author{J.\,Fink$^1$, A.\,Charnukha\altaffiliation[Present address: ]{ Department of Physics, University of California, San Diego, La Jolla, California 92093, USA}$^1$, E.D.L.\,Rienks$^2$, Z.H. Liu$^1$, S.\ Thirupathaiah$^2$\altaffiliation[Present address: ]{Solid State and  Structural Chemistry Unit, Indian Institute of Science, Banglore-560012, India},  I.\, Avigo$^3$, F.\,Roth$^4$, H.S.\,Jeevan$^5$\altaffiliation[Present address: ] {Department of Physics, PESITM, Sagar Road, 577204 Shimoga, India}, P.\,Gegenwart$^5$, M.\,Roslova$^{1,6}$, I.\,Morozov$^{1,6}$, S.\,Wurmehl$^{1,7}$, U.\,Bovensiepen$^3$, S.\,Borisenko$^1$, M.\, Vojta$^8$, B.\,B\"uchner$^{1,7}$}
\affiliation{
\mbox{$^1$Leibniz Institute for Solid State and Materials Research  Dresden, Helmholtzstr. 20, D-01069 Dresden, Germany}\\
\mbox{$^2$Helmholtz-Zentrum Berlin, Albert-Einstein-Strasse 15, D-12489 Berlin, Germany}\\
\mbox{$^3$Fakult\"at f\"ur Physik, Universit\"at Duisburg-Essen, Lotharstr. 1, D-47075 Duisburg, Germany}\\
\mbox{$^4$Center for Free-Electron Laser Science / DESY, Notkestrasse 85, D-22607 Hamburg, Germany}\\
\mbox{$^5$Institut f\"ur Physik, Universit\"at Augsburg, Universit\"atstr.1,D-86135 Augsburg, Germany}\\
\mbox{$^6$Department of Chemistry, Lomonosov Moscow State University, 119991 Moscow, Russia}\\
\mbox{$^7$Institut f\"ur Festk\"orperphysik,  Technische Universit\"at Dresden, D-01062 Dresden, Germany}\\
\mbox{$^8$ Institut f\"ur Theoretische Physik, Technische Universit\"at Dresden, D-01062 Dresden, Germany}\\}

\date{\today}

\begin{abstract}
 Angle-resolved photoemission spectroscopy (ARPES) is used to study the band dispersion and the quasiparticle scattering rates in two ferropnictides systems. Our ARPES results show  linear-in-energy dependent scattering rates  which are constant in  a wide range of control parameter and which depend on the orbital character of the bands. We demonstrate that  the linear energy dependence gives rise to weakly dispersing band with a strong mass enhancement when the band maximum crosses the chemical potential.  In   the superconducting phase
 the related  small effective Fermi energy favors a  Bardeen-Cooper-Schrieffer (BCS)\,\cite{Bardeen1957}-Bose-Einstein (BE)\,\cite{Bose1924} crossover state.
\end{abstract}


\maketitle

\paragraph{Introduction.} Unconventional/high temperature superconductivity  is observed in heavy fermion systems, cuprates,
molecular crystals, and ferropnictides close to a point in the phase diagram where, as a function of a control parameter such as pressure, chemical pressure, or doping, the antiferromagnetic order is
suppressed\,\cite{Loehneysen2007,Gegenwart2008}. A widespread view is that at this  point, which is called a quantum critical point (QCP), strong antiferromagnetic  fluctuations are a candidate for the glue mediating superconductivity and that these fluctuations would also account for the normal state non-Fermi-liquid behavior which is expressed, e.g. in a linear temperature dependence of the resistivity or to large mass enhancement of the charge carriers. In the ferropnictides\,\cite{Johnston2010} the strange normal state properties  have been observed, e.g. in transport\,\cite{Analytis2014} and thermal properties\,\cite{Meingast2012}. Theoretically a QCP in ferronpnictides has been predicted in Refs.\,\onlinecite{Dai2009,Abrahams2011}.

ARPES is a versatile method to obtain information on the energy ($\omega$) and momentum ($\mathbf{k}$) dependent  real and the imaginary part of the self-energy function $\Sigma(\omega,\mathbf{k})$ which are related to the mass enhancement and the scattering rate, respectively\,\cite{Damascelli2003}\,\footnote{see information presented in the Supplement}.

In this contribution we present ARPES results on  the band dispersion and the scattering rate as a function of the control parameter  of two ferropnictide\,\cite{Johnston2010} systems: the chemically pressurized "122" compounds \BFAP\ and \EFAP\ , and  the electron doped "111" compounds \NFCA\ and \NFRA\ .  There are numerous ARPES studies on these systems published in the literature\,\cite{He2010,Thirupathaiah2011,Liu2011a,Yoshida2011,Yi2012,Ye2014}. In the present study we have measured along three directions in the Brillouin zone, shown in Fig.\,1, to reach the high symmetry points on the three hole pockets and on the electron pockets.
\begin{figure} [tb]
  \centering
%\hspace{-1.5cm}
  \includegraphics[angle=0,width=1.1\columnwidth]{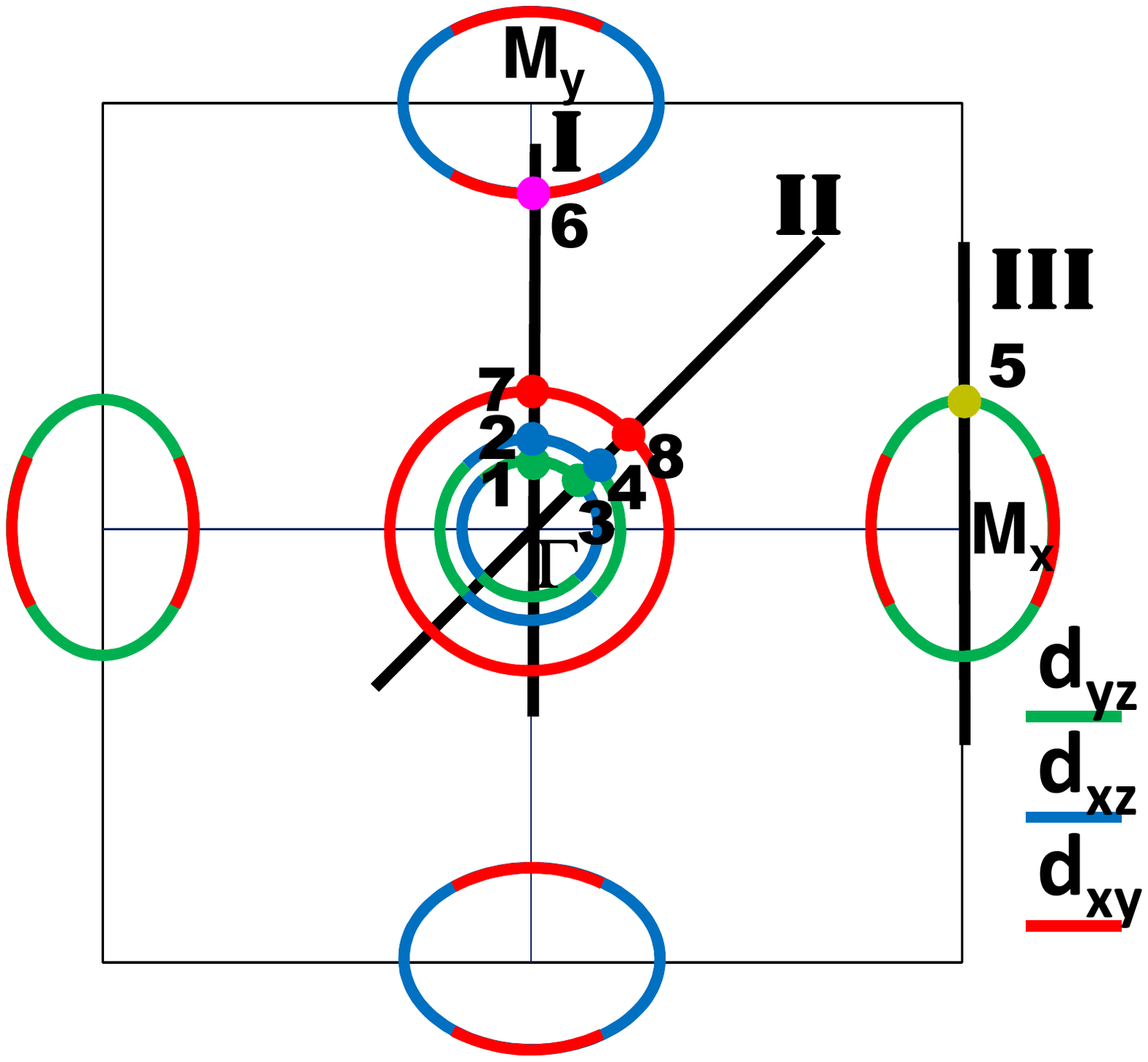}
  \caption{
(Color online) Schematic  Fermi surfaces at $k_z=0$ of ferropnictides formed by three hole pockets at $\Gamma$ and electron pockets at M. The orbital character which was adopted from Ref.\,\cite{Kemper2011} is indicated by different colors.  } 
  \label{Fermi}
 \centering
\end{figure}
In the wide range of the investigated control parameter we realize a linear-in-energy dependence of the scattering rates independent of the control parameter and no enhancement  at the supposed QCP. Furthermore we see that at optimal "doping" we observe a crossing of the top of the hole band through the Fermi level. On the basis of these  results we propose   the  following scenario: a co-action between  a  highly correlated electron liquid with a linear-in-energy scattering rate \textit{and}  a crossing of a  the top(bottom) of a hole(electron) pocket through the chemical potential causes an anomalous band dispersion at the Fermi level which leads to  a strong mass enhancement in the normal state and  to a small effective Fermi energy favoring  a Bardeen-Cooper-Schrieffer (BCS)\,\cite{Bardeen1957}-Bose-Einstein (BE)\,\cite{Bose1924} crossover state in   the superconducting phase.

\paragraph{Experimental.}Single crystals were grown using the  self-flux technique\,\cite{Jeevan2011,Steckel2014}.
ARPES measurements were conducted at the $1^2$- and $1^3$-ARPES endstations attached to the  beamline
UE112 PGM 2 at BESSY with energy and angle resolutions between 4 and 10 meV and 0.2  $^\circ$, respectively. If not otherwise stated the measuring temperature at the $1^2$- and $1^3$-ARPES endstations were 30  and 0.9 K, respectively. For the unsubstituted samples temperature above the N\'eel temperatures were used to keep the samples in the paramagnetic state. 

\paragraph{Results.}
In Fig.\,2(a) we show an energy distribution map along cut I (Fig.\,1) for the optimally "doped" \BFAP\ $x= 0.27$  measured close to the $Z=(0,0,\pi/c)$  point,   where $c$ is the lattice constant. The two inner hole pockets can be clearly resolved. Due to matrix element effects the outer hole pocket is more difficult to see in this system. In Fig.\,2(b) we show a cut at constant binding energy  $E_B=80$ meV together with a fit to four Lorentzians. 
\begin{figure}[tb]
 \centering
\includegraphics[angle=0,width=\columnwidth]{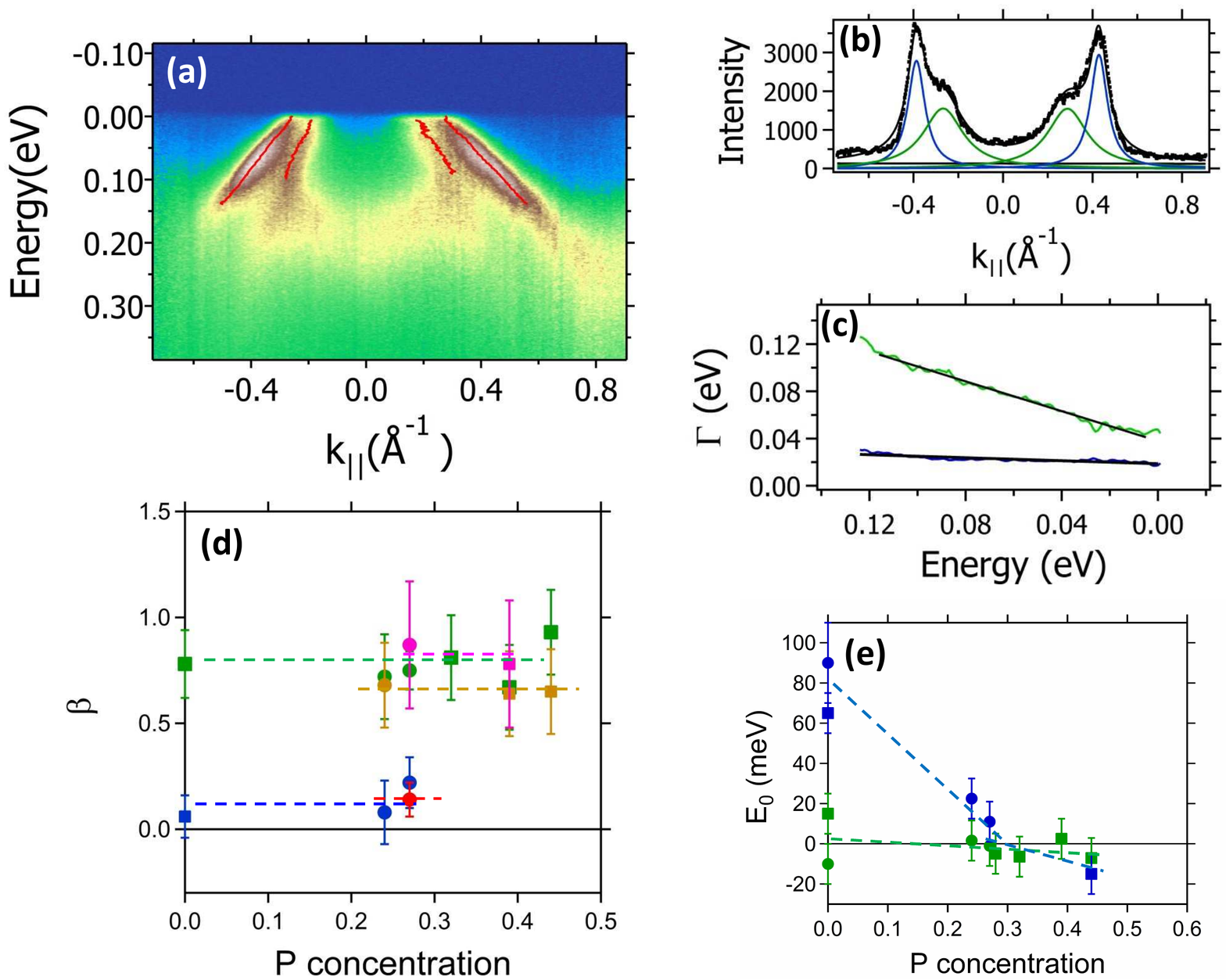}
%\hspace{-5cm}
  \caption{
  (Color online) ARPES data  \BFAP\ (circles) and \EFAP\ (squares) (a) Energy distribution map along cut I near $Z$ of \BFAP\  $x=0.27$.
 The red lines depict dispersions derived from fits. (b) Momentum distribution curve taken from the data in (a) at the binding energy of 80 meV. (c) Scattering rates $\Gamma$ as a function of energy  for the inner  and the middle  hole pocket together with a linear fit (black). (d) Compilation of all $\beta$ values for various high symmetry points for $k_z=0$ as a function of the P concentration. 
 (e) Top of the inner (green) and the middle (blue) hole pocket as a function of the P concentration. The color code corresponds to that of the high symmetry points in Fig.\,1. The  dashed lines are guides to the eye.}
  \label{phos}
 \centering
\end{figure} 
 From those fits the life-time broadening  $\Gamma$ as a function of the energy is derived  as presented in Fig.\,2(c). For both hole pockets, we find a non-Fermi-liquid linear-in-energy dependence  in a large energy range of $5 \leq \omega \leq120$ meV, which can be described by $\Gamma=\alpha +\beta\omega$. While  $\alpha$ is determined by elastic scattering,  $\beta$ is determined by inelastic scattering processes. $\beta$ is much larger for the inner hole pocket compared to the value of the middle hole pocket. Similar results have been obtained previously for the system \BFCA\  \cite{Brouet2011}. 
In Fig.\,2(d) we present a compilation of all available data of $\beta$  for \BFAP\ and \EFAP\ taken at $k_z=0$. 
 
In Fig.2(e) we show the top of the middle and the inner hole pocket as derived from a parabolic fit to the measured dispersion. In the unsubstituted samples a splitting between $xz$ and the $yz$ states of the order of 50 meV is observed  in the tetragonal paramagnetic phase which is probably caused by nematic fluctuations\,\cite{Yi2011,Fisher2011}. As expected from other studies\,\cite{Fisher2011} the splitting decreases and disappears near optimal "doping". While we observe  a decrease of the Fermi wave vector at $\Gamma$ with increasing "doping" concentration, at  $Z$   the Fermi wave vector increases.  Thus we observe a Lifshitz transition of a Fermi  cylinder at low P concentrations to an ellipsoid around the $Z$ point at higher P concentrations. This Lifshitz transition  has been detected in  several ARPES studies of 122 systems\,\cite{Brouet2009,Thirupathaiah2011}.
In the present ARPES study, within error bars, we have not found   any major differences in the electronic structure between  \BFAP\  and \EFAP\   . 

The 111 compounds are ideal for ARPES experiments since they do not form a polar surface after cleaving and therefore,  one is certain to find bulk properties\,\cite{Lankau2010}. On the other hand, they have the disadvantage that close to the Fermi level the middle hole pocket  hybridizes with the inner hole pocket.  As a consequence, reliable results on the scattering rates can be only derived at higher binding energies.
In Fig.\,3(a) we show an  energy distribution map of  the optimally doped \NFRA\ $x= 0.027$   close to the $\Gamma$  point along cut I. All three hole pockets can be clearly resolved. In Fig.\,3(b) we depict a cut at constant $E_B=40$ meV together with a fit  by six Lorentzians. 
\begin{figure}[tb]
  \centering
\hspace{-1.4cm}
  \includegraphics[angle=0,width=\columnwidth]{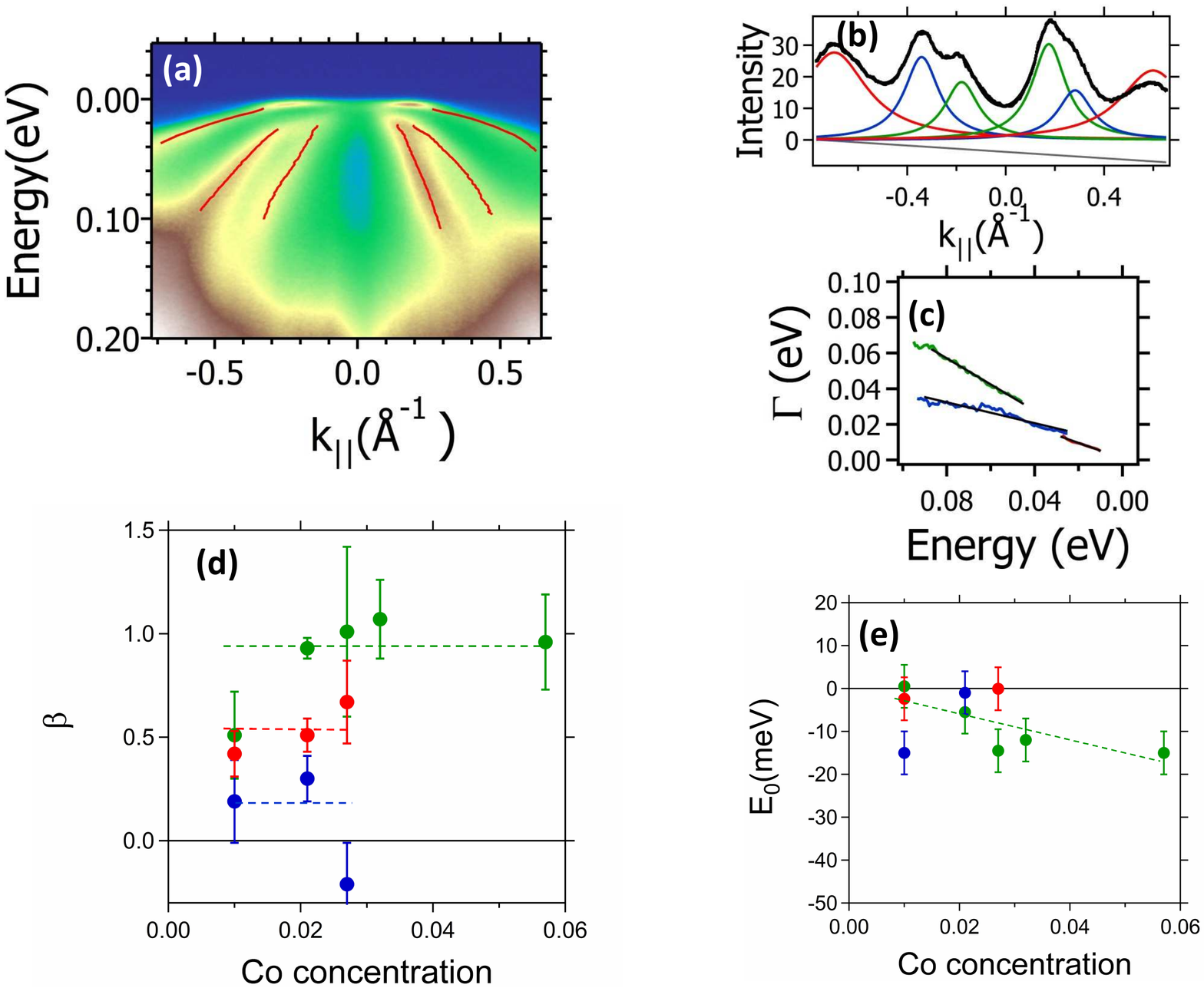}
  \caption{
(Color online) ARPES data of 111 compounds.
(a) Energy distribution map along cut I near $\Gamma$ of \NFRA\  $x=0.027$. The red lines correspond to dispersions derived from fits. (b) Momentum distribution curve taken from the data in (a) at $E_B=40$ meV together with a fit. (c) Scattering rates $\Gamma$ as a function of energy for the three hole pockets together with a linear fit (black). (d) Compilation of all $\beta$ values of \NFCA\ for various high symmetry points for $k_z=0$  as a function of the Co concentration.  
 (e) Top of the inner , the middle , and the outer hole pocket as a function of the Co concentration. The color code corresponds to that of the high symmetry points in Fig.\,1. The horizontal  dashed lines are guides to the eye.} 
 \label{cob}
 \centering
\end{figure}

 From those fits  we derive   $\Gamma$ as a function of the energy   as presented in Fig.\,3(c). Again for all three hole pockets, $\Gamma$  increases  linearly with increasing energy  in the range where there is no band overlapping.  For the inner, middle, and outer hole pockets we derive $\beta$ values of 1.0, $\approx$0, and 0.69, respectively. Comparing the superconducting gaps $\Delta$ for the similar compound  \LFA\  \,\cite{Borisenko2012,Umezawa2012} with the present values of $\Gamma$ shows  that the strength of the scattering rate is directly  related to the size of the superconducting gap.
 In Fig.\,3(d) we present a compilation of all $\beta$ values for \NFCA\  near $\Gamma$  as a function of Co concentration. As in the case of the 122 compounds, the $\beta$
values are rather independent of the control parameter and no enhancement  at optimal doping ($x=0.027$) is observed.  Finally, in Fig.\,3(e) the top of the hole pockets as a function of the Co concentration is depicted. As in the case of the 122 compounds, the crossing of the top of the inner hole pocket, having the strongest scattering rate, occurs very close to optimal doping.

For all four compounds we have obtained similar results at the other high-symmetry points: the $\beta$ values are not dependent on the control parameter and there is  no $k_z$ dependence\,\cite{Note1}.

\paragraph{Discussion.}Our experimental results can be compared with those from theoretical calculations in the framework of density functional theory  (DFT) combined with dynamical mean-field theory (DMFT). Interestingly, non-Fermi-liquid-like self-energies have been reported\,\cite{Werner2012,Yin2012}  which could be fitted by sub-linear power laws over a range of elevated energies. This incoherent-metal behavior has been attributed to an interplay of Hubbard and Hund's-rule couplings in a multi-band system: there is a sizable regime where carriers are  strongly scattered by slowly fluctuating unquenched spins. Importantly, this source of scattering is essentially local in space, implying that it is weakly influenced by longer-range spin correlations and can thus persist over an extended part of the phase diagram. We believe that our observation of a strong, weakly doping-dependent, and approximately energy-linear single-particle scattering rate is consistent with originating from local correlation physics as the one described by DMFT. We point out, however, that the doping dependence of the scattering rate as reported in Ref.\,\cite{Werner2012}, where non-Fermi-liquid behavior has been seen only for hole but not for electron doping, appears inconsistent with our results.

In the highly correlated systems inelastic scattering processes  are determined by intra (having the same orbital character) and inter-band (having different orbital character) transitions. The difference in the scattering rates between the hole pockets have been predicted by theoretical calculations\,\cite{Graser2009,Kemper2011} as the scattering by intra-band transitions should be larger than that by inter-band transitions. 

 From the scattering rates we have calculated  $\Im\Sigma\propto\Gamma$\,\cite{Damascelli2003,Note1}. Making the assumption that $\Im\Sigma$ is also linear in energy at low energies and constant above a ultraviolet cutoff energy $\omega_c$ , also used in the phenomenological  marginal Fermi liquid theory\,\cite{Varma2002}, one can define a coupling constant $\lambda_{MF}$  by the relation  $\Im\Sigma_{MF}=\frac{\pi}{2}\lambda_{MF}u$, where $u=max(|\omega|,kT)$ and  $kT$ is the thermal energy. 
The comparison with experimental data near point 1 for \BFAP\ yields a value of $\lambda_{MF}=1.5$ which is slightly larger than values derived for  cuprates\,\cite{Valla1999a}. Further values for $\lambda_{MF}$ are presented in the Supplement\,\cite{Note1}. 
 
\begin{figure}[tb]
  \centering
%\hspace{7cm}
  \includegraphics[angle=0,width=\columnwidth]{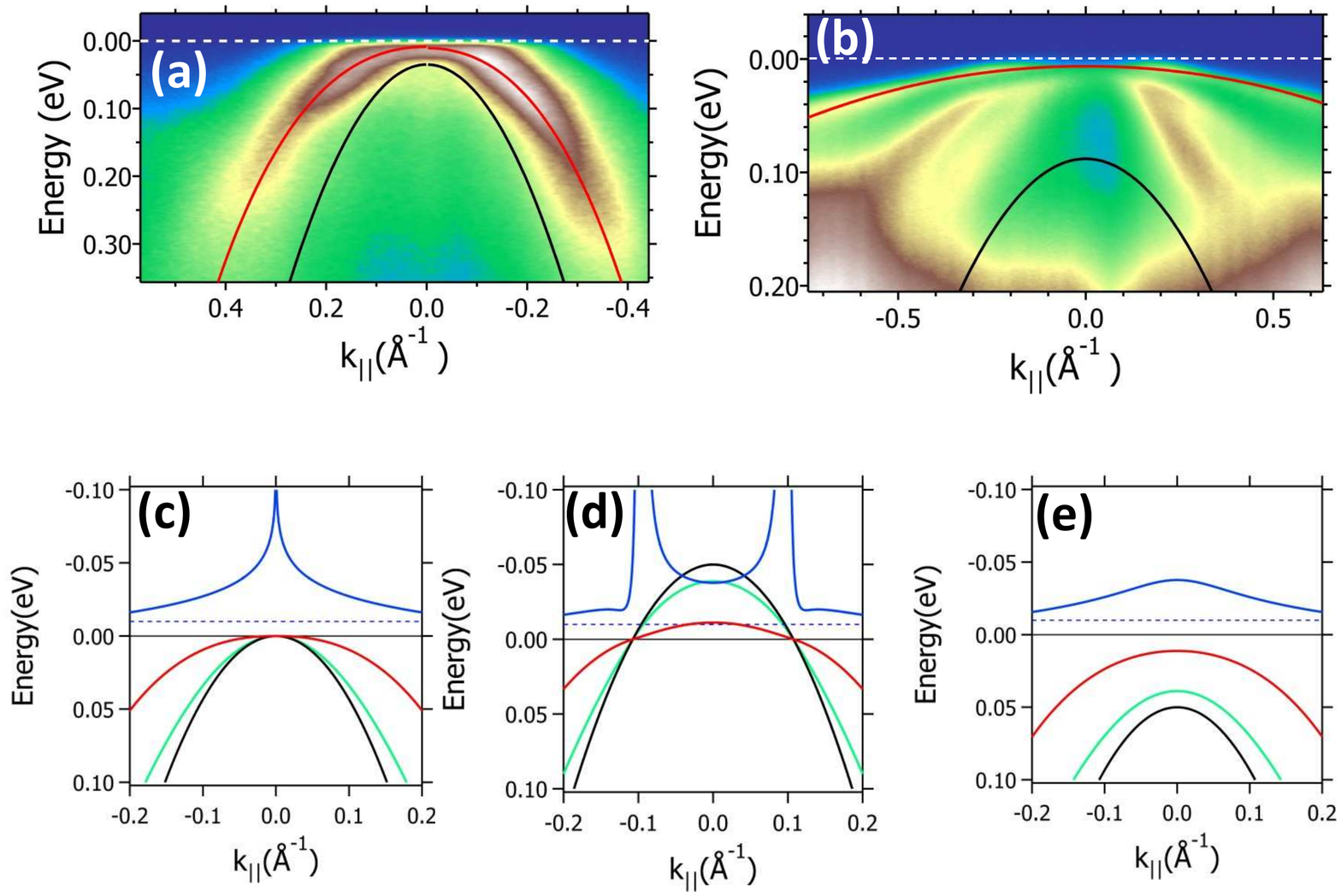}
  \caption{
  (Color online) (a) Energy distribution map along cut I near $\Gamma$ of \BFAP\ $x=0.24$ together with a fit (red line)   and the bare particle dispersion (black line). (b)   Energy distribution map along cut I near $\Gamma$  of \NFRA\ $x=0.027$ together with the fitted dispersion for the outer hole pocket (red line)  and the bare particle dispersion ( black line). (c)  Bare particle dispersion (black line) with the top of the hole pocket at  the Fermi level   together with calculations of $\Re\Sigma_{MF}$ (green line) and the renormalized dispersion (red line). In addition we show the effective mass $m^*/m_0$ (blue line) and the $m^*/m_0=1$ line (thin dashed blue line).    (d) The same as  (c)  but with the top of the hole pocket 50 meV above the Fermi level. (e) The same as  (c)  but with the top of the hole pocket 50 meV below the Fermi level.} 
 \label{mass}
 \centering
\end{figure}
 Since $\Re\Sigma$ and $\Im\Sigma$ are connected by the Kramers-Kronig relation, we  also use a non-Fermi-liquid $\Re\Sigma$ which can be derived from a numerical calculation or from the analytic expression $\Re\Sigma_{MF}=\frac{1}{2}\lambda_{MF}\omega ln\frac{\omega_c}{u}$\,\cite{Varma2002}.  
 In Fig.\,4(a) we show an energy distribution map of \BFAP\ $x=0.24$ measured near $\Gamma$ along cut I using a vertical photon polarization.  This dispersion could be fitted  combining a parabolic  bare particle dispersion  derived from DFT calculations\,\cite{Singh2008} and $\Re\Sigma_{MF}$. Choosing a cut-of energy $\omega_c= 1.5 $ eV  we obtain a $\lambda_{MF}= 1.4$ which is close to the value derived from the scattering rate. 
 Analogous data for \NFRA\  are shown in Fig.\,4(b). Here we have fitted the dispersion of the outer band (red line) using $\Re\Sigma_{MF}$.  A coupling constant $\lambda_{MF}=1.2$ is obtained. In Fig.\,4(a) and (b) we  obtain a dispersion which at the Fermi level is strongly reduced compared to the bare particle dispersion. We emphasize that the  weak dispersion near the Fermi level in \NFRA\ is not caused by a coherence peak in the superconducting state since the measurement presented in Fig.\,4(b) was performed in the normal state at $T= 26$ K.

Using $\Re\Sigma_{MF}$, we can now calculate the renormalized dispersion close to the Fermi level as a function of the band position. We used the parameters typical of the inner hole pocket of \BFAP\ : $\lambda_{MF}=1.5$, $\omega_c=1.5 $ eV, and a bare particle dispersion from DFT calculations on  $\mathrm{BaFe}_{2}\mathrm{As}_{2}$\,\cite{Singh2008}. In Fig.\,4(c) we show such  calculations for a bare particle hole pocket touching the chemical potential , for $\Re\Sigma $, and for the renormalized band. Due to the fact that $\Re\Sigma $ scales to zero logarithmically and is very close to the bare particle dispersion, there is a strong renormalization close to the Fermi level which leads to a weakly dispersive band. The extension of this singularity increases with increasing $\lambda_{MF}$ and increasing $m_0$.  The mass enhancement  is energy dependent and diverges at zero energy. In Fig.\,4(d) we depict an analogous  calculation but with the top of the bare hole pocket shifted 50 meV above the Fermi level. At the $\Gamma$ point the calculated  dispersion shows a renormalization by a factor of four. One immediately realizes that the extension of the weakly  renormalized dispersion at the Fermi level  is strongly reduced.  Moreover, the divergence of the effective mass curve is shifted by $ E  \approx 2.5$  meV into the unoccupied states. Below  and at the Fermi level there is only a moderate mass enhancement for the occupied electronic states. We emphasize that in this case, where the top of the hole pocket is not at the Fermi level, there is no divergent mass enhancement. This means that the strong low-energy scattering rates  of a moderately correlated system alone cannot describe the strange normal state properties at a specific value of the control parameter.  In Fig.\,4(e) we present analogous calculations for a bare particle hole pocket shifted below 50 meV below the Fermi level. At $\Gamma$ the bare particle dispersion is renormalized by a factor $\approx$ 5 and the mass enhancement  is strongly reduced.
Analogous calculations, in which $\Im\Sigma$ is extrapolated below 5 meV by a Fermi liquid behavior  ($\Im\Sigma \propto \omega ^2$) yield also a large mass enhancement at the point where the hole pocket crosses the Fermi level\,\cite{Note1}.

These calculations  show that
the crossing of  the top of the hole pocket having there a constant density of states (DOS) typical of  a two-dimensional electronic structure  plays an important role for  these correlated systems.  It can  explain  the divergent mass enhancement in the transport properties observed in \BFAP\ near optimal "doping"\,\cite{Analytis2014} and the strange behavior of  the thermal properties of \BFCA\ \,\cite{Meingast2012}.

There are several ARPES publications on iron-based  superconductors  which show clear indications of weakly dispersing bands at the Fermi level, which cannot be explained on the basis of DFT calculations. In a recent ARPES study of
 ferrochalcogenides  an apparently non-dispersive peak at  the chemical potential  has been detected  which is  not in agreement with DFT calculations\,\cite{Starowicz2013}. Moreover, in the same system  "shallow pockets" have been detected which could lead to a  BCS-BE crossover in the superconducting phase\,\cite{Lubashevsky2012}. These anomalous band dispersions can be well described by the scenario presented here.

Different from the iron-based superconductors, in the cuprates there is already in the bare particle band structure a peaked DOS  due to a saddle point in the band structure. There is a longstanding discussion on  the importance of this van Hove singularity,  which crosses the Fermi level near optimal doping\,\cite{Markiewicz1997,Kastrinakis1999}  and on a breakdown of the Migdal theorem\,\cite{Migdal1958}  which was the basis of the standard BCS theory of superconductivity. This breakdown requires the inclusion of non-adiabatic effects and the generalization of the Eliashberg equations\cite{Grimaldi1995}  interpolating between the BCS theory and the BE condensation\,\cite{Andrenacci1999}.   In an ARPES study on \ybco\ an "extended" van Hove singularity has been found\,\cite{Gofron1994} which nicely can be explained by the scenario described above: a co-action between a highly correlated electron liquid behavior and a weakly dispersing band (the saddle point) moving through the Fermi level.

\paragraph{Summary.}We have measured the scattering rates of two ferropnictide systems at various high symmetry points. In a wide range of the  control parameter  we observe constant linear-in-energy  scattering rates.  In particular, we see no enhancement  of the scattering rates close to the expected QCP.  Furthermore at optimal "doping" the top of the hole pockets having the  largest scattering rates crosses the Fermi level. The strange normal state transport and thermal properties near optimal "doping" and probably also the unconventional superconductivity can be related to  a proposed scenario in which a co-action of the real part of a non- Fermi liquid self-energy with an edge in the DOS  at the Fermi level leads to a weakly dispersing band at the chemical potential and to  strongly enhanced  effective masses.  The results can be generalized to other unconventional superconductors and possibly are a recipe  for future search of new high-$T_c$ superconductors.

\bibliographystyle{phaip}
\bibliography{Pnictide}

\end{document}

\title{Supplementary material for: Non-Fermi-liquid scattering rates and anomalous band dispersion in ferropnictides}
%\author{J. Fink}
%\affiliation{Leibniz Institute for Solid State and Materials Research  Dresden, Helmholtzstr. 20, D-01069 Dresden, Germany}
%\author{A. Charnukha}
%\affiliation{Leibniz Institute for Solid State and Materials Research  Dresden,\,Helmholtzstr. 20,~D-01069 Dresden,~Germany}
%\author{E.D.L. Rienks}
%\affiliation{\nobreak{Helmholtz-Zentrum Berlin,~Albert-Einstein-Strasse 15,~D-12489 Berlin,~Germany}}
%\author{S. Thirupathaiah}
%\affiliation{Helmholtz-Zentrum Berlin, Albert-Einstein-Strasse 15, D-12489 Berlin, Germany}
%\author{I. Avigo}
%\affiliation{Fakult\"at f\"ur Physik, Universit\"at Duisburg-Essen, Lotharstr. 1, D-47075 Duisburg, Germany}
\author{J.\,Fink$^1$, A.\,Charnukha\altaffiliation[Present address: ]{ Department of Physics, University of California, San Diego, La Jolla, California 92093, USA}$^1$, E.D.L.\,Rienks$^2$, Z.H. Liu$^1$, S.\ Thirupathaiah$^2$\altaffiliation[Present address: ]{Solid State and  Structural Chemistry Unit, Indian Institute of Science, Banglore-560012, India},  I.\, Avigo$^3$, F.\,Roth$^4$, H.S.\,Jeevan$^5$\altaffiliation[Present address: ] {Department of Physics, PESITM, Sagar Road, 577204 Shimoga, India}, P.\,Gegenwart$^5$, M.\,Roslova$^{1,6}$, I.\,Morozov$^{1,6}$, S.\,Wurmehl$^{1,7}$, U.\,Bovensiepen$^3$, S.\,Borisenko$^1$, M.\, Vojta$^8$, B.\,B\"uchner$^{1,7}$}
\affiliation{
\mbox{$^1$Leibniz Institute for Solid State and Materials Research  Dresden, Helmholtzstr. 20, D-01069 Dresden, Germany}\\
\mbox{$^2$Helmholtz-Zentrum Berlin, Albert-Einstein-Strasse 15, D-12489 Berlin, Germany}\\
\mbox{$^3$Fakult\"at f\"ur Physik, Universit\"at Duisburg-Essen, Lotharstr. 1, D-47075 Duisburg, Germany}\\
\mbox{$^4$Center for Free-Electron Laser Science / DESY, Notkestrasse 85, D-22607 Hamburg, Germany}\\
\mbox{$^5$Institut f\"ur Physik, Universit\"at Augsburg, Universit\"atstr.1,D-86135 Augsburg, Germany}\\
\mbox{$^6$Department of Chemistry, Lomonosov Moscow State University, 119991 Moscow, Russia}\\
\mbox{$^7$Institut f\"ur Festk\"orperphysik,  Technische Universit\"at Dresden, D-01062 Dresden, Germany}\\
\mbox{$^8$ Institut f\"ur Theoretische Physik, Technische Universit\"at Dresden, D-01062 Dresden, Germany}\\}

\date{\today}

\begin{abstract}
 In this Supplement we provide additional information on the scattering rate of the quasiparticles in the "122" compounds \BFAP\ and \EFAP\ and the "111" compounds \NFCA\ and \NFRA\ near the high-symmetry points in the Brillouin zone. Furthermore we present additional calculations on the band dispersion and the effective mass in a model where the scattering rates at low energy obey a Fermi liquid behavior and at higher energy a marginal Fermi liquid behavior.

\end{abstract}

\pacs{74.25.Jb, 74.72.Ek, 79.60.−i }

\maketitle

\paragraph{Spectral function.}

ARPES measures the energy ($\omega$) and momentum ($\mathbf{k}$) dependent spectral function $A(\omega ,\mathbf{k})$  multiplied by a transition matrix element\,\cite{Damascelli2003}. The former is given by
\begin{eqnarray}
A(\omega ,\mathbf{k})&=&-\frac{1}{\pi}\frac{\Im\Sigma(\omega ,\mathbf{k})}
{[\omega -\epsilon _{\mathbf{k}}-\Re\Sigma(\omega ,\mathbf{k})]^2+[\Im\Sigma(\omega ,\mathbf{k})]^2},
\label{26_1}
\end{eqnarray}
where $\epsilon _{\mathbf{k}}$  is the bare particle energy, $\Re\Sigma(\omega ,\mathbf{k})$ is the real part  and $\Im\Sigma(\omega ,\mathbf{k} ) $ is the imaginary part of the self-energy. For small $\Im\Sigma(\omega ,\mathbf{k})$ the maxima of the spectral function (the dispersion) are determined by the equation
\begin{eqnarray}
\omega -\epsilon _{\mathbf{k}}-\Re\Sigma(\omega ,\mathbf{k})=0.
\end{eqnarray}
In the case when $\Re\Sigma(\omega ,\mathbf{k})=-\lambda \omega$  as in a Fermi liquid close to the chemical potential and when $\Re\Sigma(\omega ,\mathbf{k})$ is independent of the momentum, the dispersion follows $\omega=\epsilon _{\mathbf{k}}/(1+\lambda)$, i.e.,  with increasing coupling constant $\lambda$ the maximum of the spectral function is shifted to lower binding energy which means the effective mass is enhanced.  In  a marginal Fermi liquid,  a concept originally devised to phenomenologically describe the normal state of the  cuprate superconductors, one assumes that   the  scattering rates at low energy are much higher compared to a normal Fermi liquid due to strong correlation effects and/or nesting. The self-energy is given by\,\cite{Varma2002} 
\begin{eqnarray}
\Sigma=\frac{1}{2}\lambda_{MF}\omega ln\frac{\omega_c}{u}-i\frac{\pi}{2}\lambda_{MF}u,
\end{eqnarray}
where $u=max(|\omega|,kT)$,  $kT$ is the thermal energy,  $\lambda_{MF}$ is a coupling constant, and $\omega_c$ is an ultraviolet cutoff energy.  In this case $\Re\Sigma$  is
no longer  linear in energy and one has to solve equation (2)  numerically to derive the dispersion.

From the measured width W at constant energy  one can determine the life time broadening or the scattering rate of the quasi-particles  $\Gamma=Wv^*$ where $v^*$ is the renormalized velocity. The imaginary part of the self-energy is then given by the relation  
 $\Im\Sigma=\Gamma\frac{ v_k}{v^*_k}=\Gamma\frac{m^*}{m_0}$,
where $v_k$  is  the bare velocity,  $m_0$ is the bare band mass, and $m^*$ is the renormalized mass\,\cite{Damascelli2003}.  In the present study $m_0$ was  taken from DFT calculations of the "undoped" compounds\,\cite{Singh2008,Deng2009} while $m^*$ values were determined from a  fit of the measured dispersion.
\begin{figure} [!htbp]
  \centering
%\hspace{-1.5cm}
  \includegraphics[angle=0,width=1.0\columnwidth]{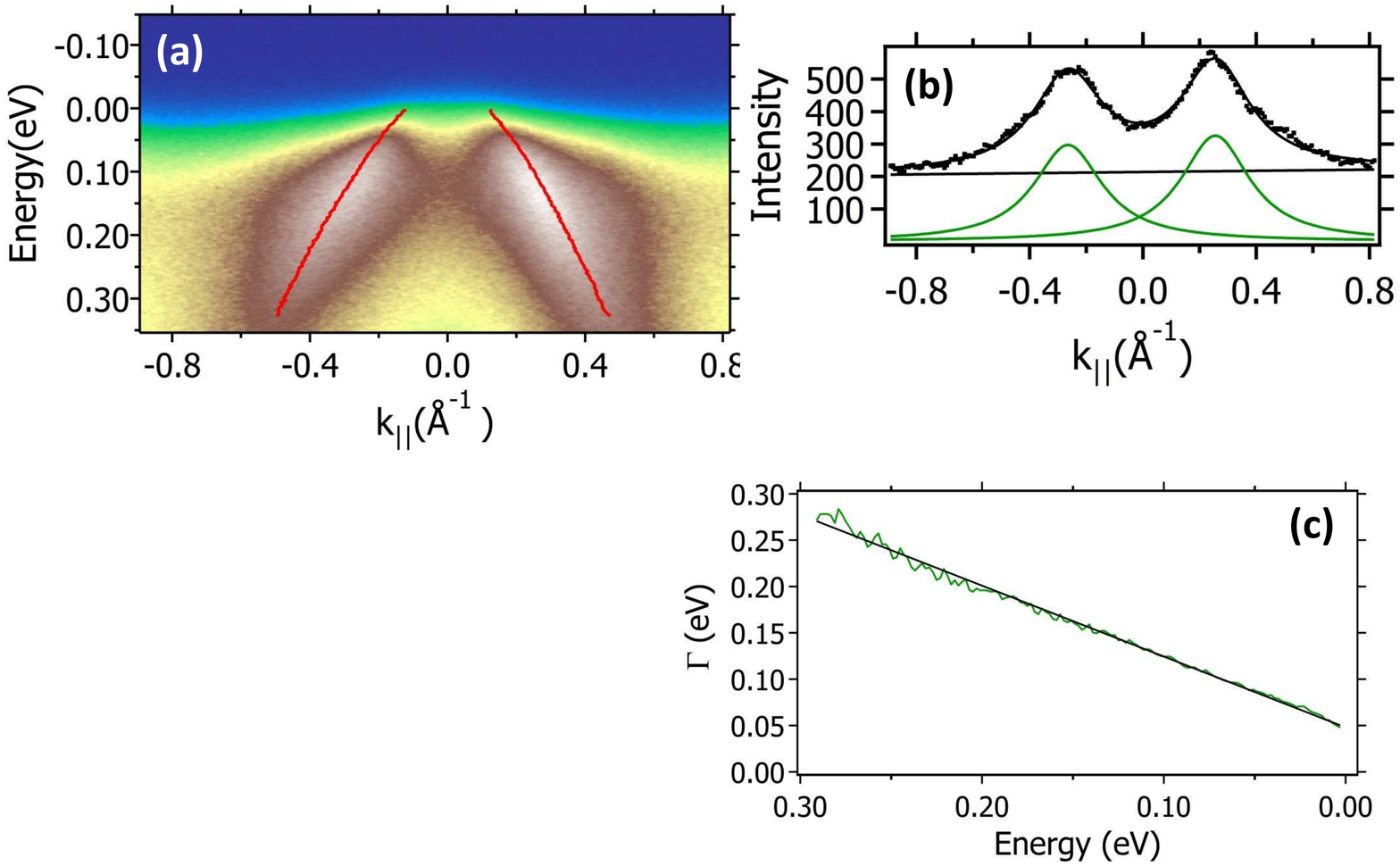}
\vspace{-1cm}
  \caption{(Color online) ARPES data of  EuFe$_2$As$_2$ recorded at a temperature $T=200$ K with vertically polarized  photons having an energy h$\nu$=96 eV.
(a) Energy distribution map along cut I (see Fig.\,1 of the main paper) near $\Gamma$. The red lines correspond to dispersions derived from fits. (b) Momentum distribution curve taken from the data in (a) at a binding energy  of 100 meV together with a fit to two Lorentzians. (c) Scattering rate $\Gamma$ at point 1 (see Fig.\,1 of the main paper) as a function of energy (green)
for the inner hole pocket together with a linear fit (black). } 
  \label{Fermi}
 \centering
\end{figure}
\paragraph{Additional data of scattering rates.}

 In Fig.\,1(a) we show an energy distribution map  of of the inner hole pocket near $\Gamma$ of undoped EuFe$_2$As$_2$. In Fig.\,2(b) we show a momentum distribution curve (MDC) taken from the data in (a) at  a binding energy  $E_B=$ 100 meV together with a fit to two  Lorentzians.  From the width of the Lorentzians and from the bare particle velocity taken from density functional theory (DFT) calculations we obtain the scattering rates $\Gamma$ as a function of binding energy. A fit clearly shows that also in the undoped pnictides the scattering rate of the quasi-particles is linear in energy over a large energy range.
\begin{figure}[!tbp]
 \centering
\includegraphics[angle=0,width=\columnwidth]{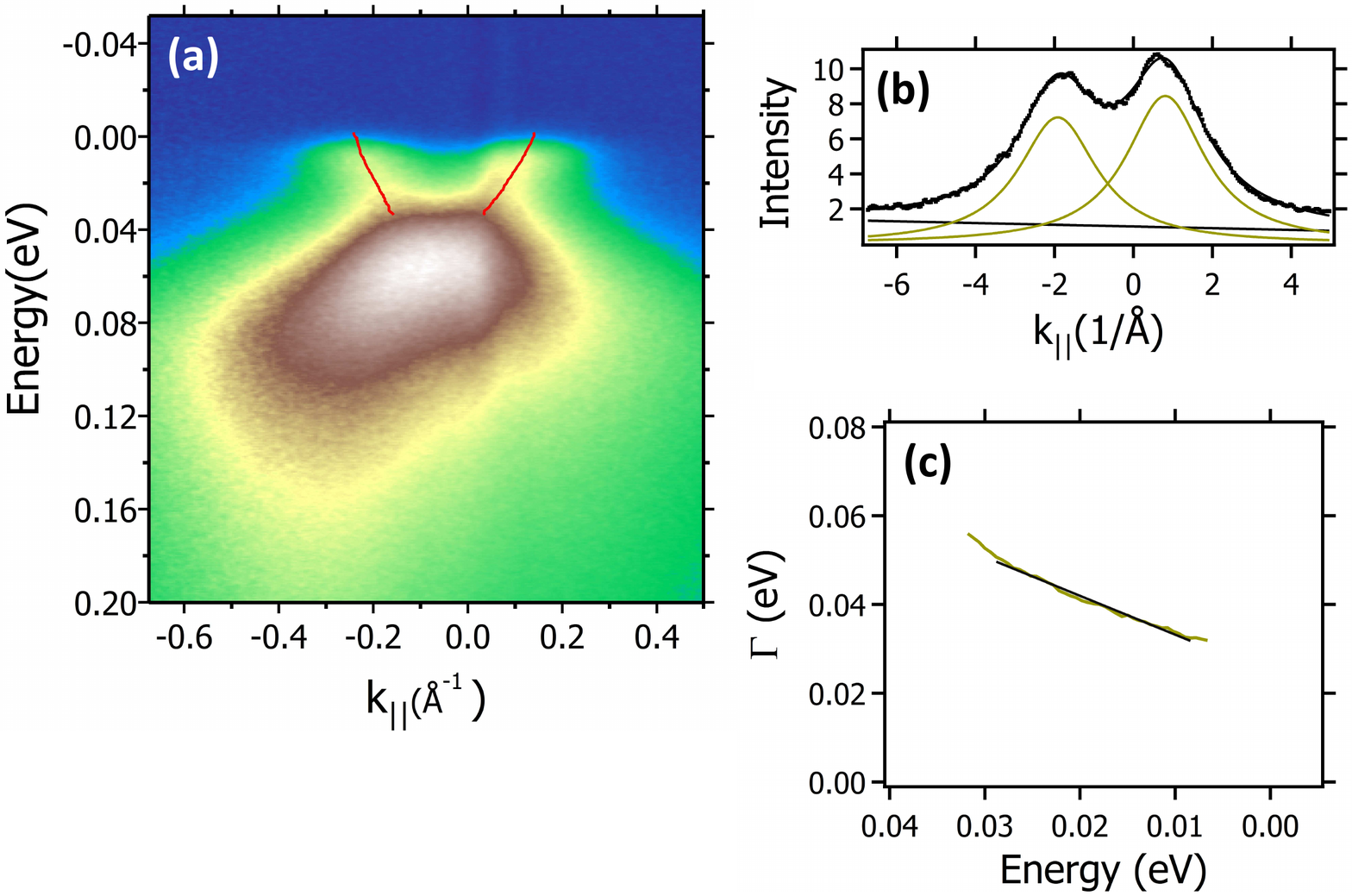}
%\hspace{-5cm}
  \caption{
  (Color online) ARPES data  of overdoped \NFCA\ $x=$0.057 recorded at $T=$1 K with vertically polarized photons having an energy of h$\nu$=58 eV. (a) Energy distribution map along cut III (see Fig.\,1 of the main paper) near $A$ ($\frac{\pi}{a},0,\frac{\pi}{c}$) where $a$ and $c$ are lattice constants.
 The red lines depict dispersions derived from fits. (b) Momentum distribution curve taken from the data in (a) at a binding energy of 20 meV together with a fit. (c) Scattering rates $\Gamma$ at point 5 (see Fig.\,1 of the main paper) and $k_z=\frac{\pi}{c}$ as a function of energy for the  hole pocket (yellow-green) together with a linear fit (black).}
  \label{phos}
 \centering
\end{figure}
 
In Fig.\,2 we show analogous data for the electron pocket at $A$=$(\frac{\pi}{a},0,\frac{\pi}{c})$ for  \NFCA\ $x=$0.057. Also for the electron pocket in this overdoped pnictide a liner-in-energy scattering rate is observed in a wide energy range. We remark that the evaluation of the scattering rates of the electron pockets is only feasible for overdoped compounds, since for underdoped systems the bottom of the electron pockets is very close to the Fermi level.
\begin{figure}[!tbp]
  \centering
\hspace{-1.4cm}
  \includegraphics[angle=0,width=\columnwidth]{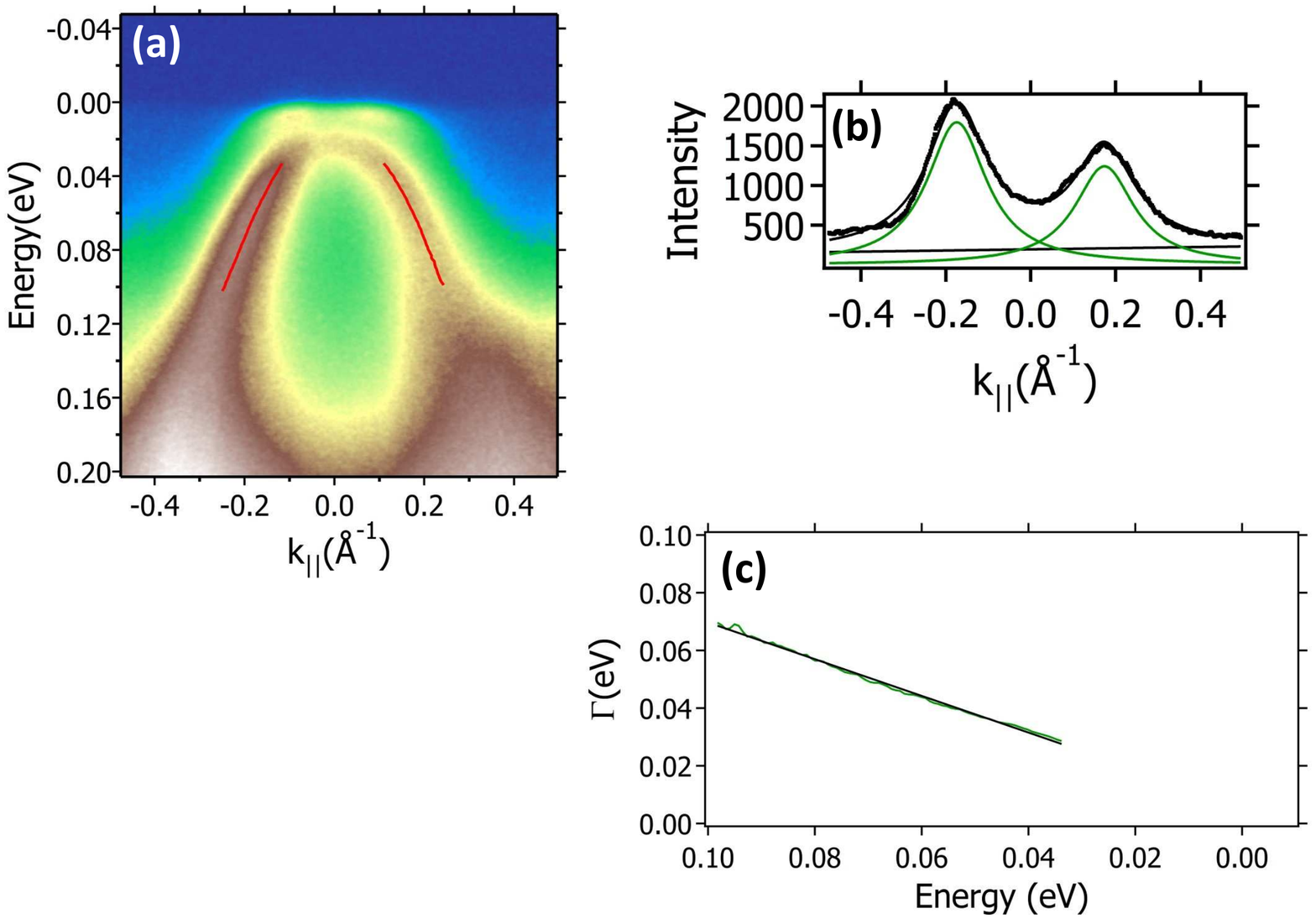}
  \caption{(Color online) ARPES data  of overdoped \NFCA\ $x=$0.057 recorded at $T=$1K with horizontally polarized photons having an energy of h$\nu$=41 eV.  (a) Energy distribution map along cut I (see Fig.\,1 of the main paper) near $\Gamma$.
 The red lines depict dispersions derived from fits. (b) Momentum distribution curve taken from the data in (a) at a binding energy of 60 meV. (c) Scattering rate $\Gamma$ at point 1 (see Fig.\,1 of the main paper) as a function of energy (green)
for the inner hole pocket together with a linear fit (black).} 
 \label{cob}
 \centering
\end{figure}

In Fig.\,3 we show analogous data for the inner hole pocket of overdoped \NFCA\ $x=$0.057. Using horizontally polarized photons the matrix element is large only for the inner hole pocket. Also in this case a linear-in-energy scattering rate is derived, as shown in Fig.\,3(c). From the fit we obtain for the slope of the scattering rate $\beta=$0.62. 
\begin{figure}[!tbp]
  \centering
  \includegraphics[angle=0,width=\columnwidth]{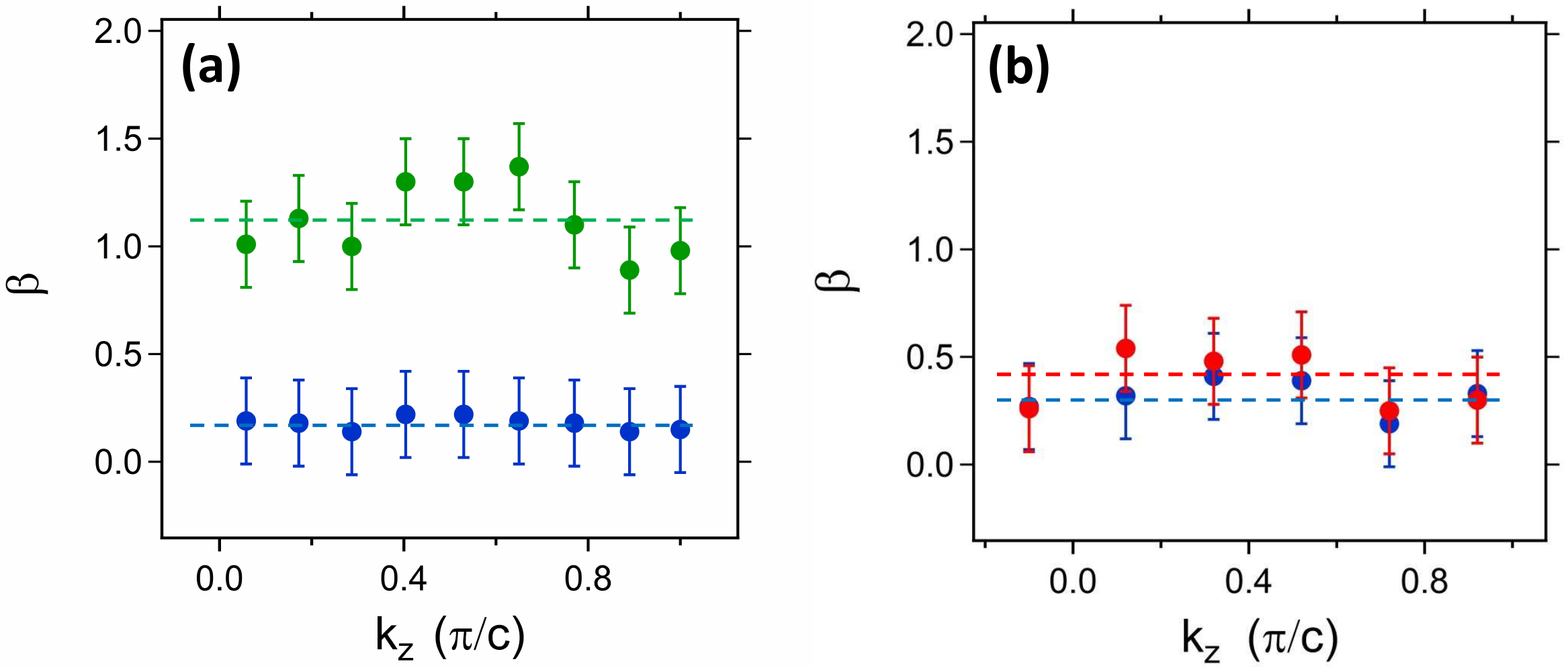}
\vspace{-2.5cm}
  \caption{
  (Color online) $k_z$-dependence of the slopes $\beta$ of the scattering rates of quasi-particles in ferropnictides for cut I (see Fig.\,1 of the main paper) and $k_z=0$. (a) \BFAP\ $x=0.27$, inner hole pocket (green) and  middle hole pocket (blue) (b) \NFRA\ $x$=0.027, middle hole pocket (blue and outer hole pocket (red). The dashed lines are guides to the eye.} 
 \label{mass}
 \centering
\end{figure}

In Fig.\,4 we show the $k_z$ dependence of the scattering rates of quasi-particles in ferropnictides. In Fig.\,4(a) we present data of optimally doped  \BFAP\ $x=0.27$ of the inner and the middle hole pocket. In Fig.\,4(b) we depict analogous data of the middle and the outer hole pockets in \NFRA\ $x$=0.027.  Although there are several reports claiming a change of the orbital character of the hole pockets as a function of $k_z$ \cite{Malaeb2009,Thirupathaiah2010,Yoshida2011},  in both cases, within error bars, there is no change of the scattering rates as a function of $k_z$.
\begin{figure}[!tbp]
  \centering
  \includegraphics[angle=0,width=\columnwidth]{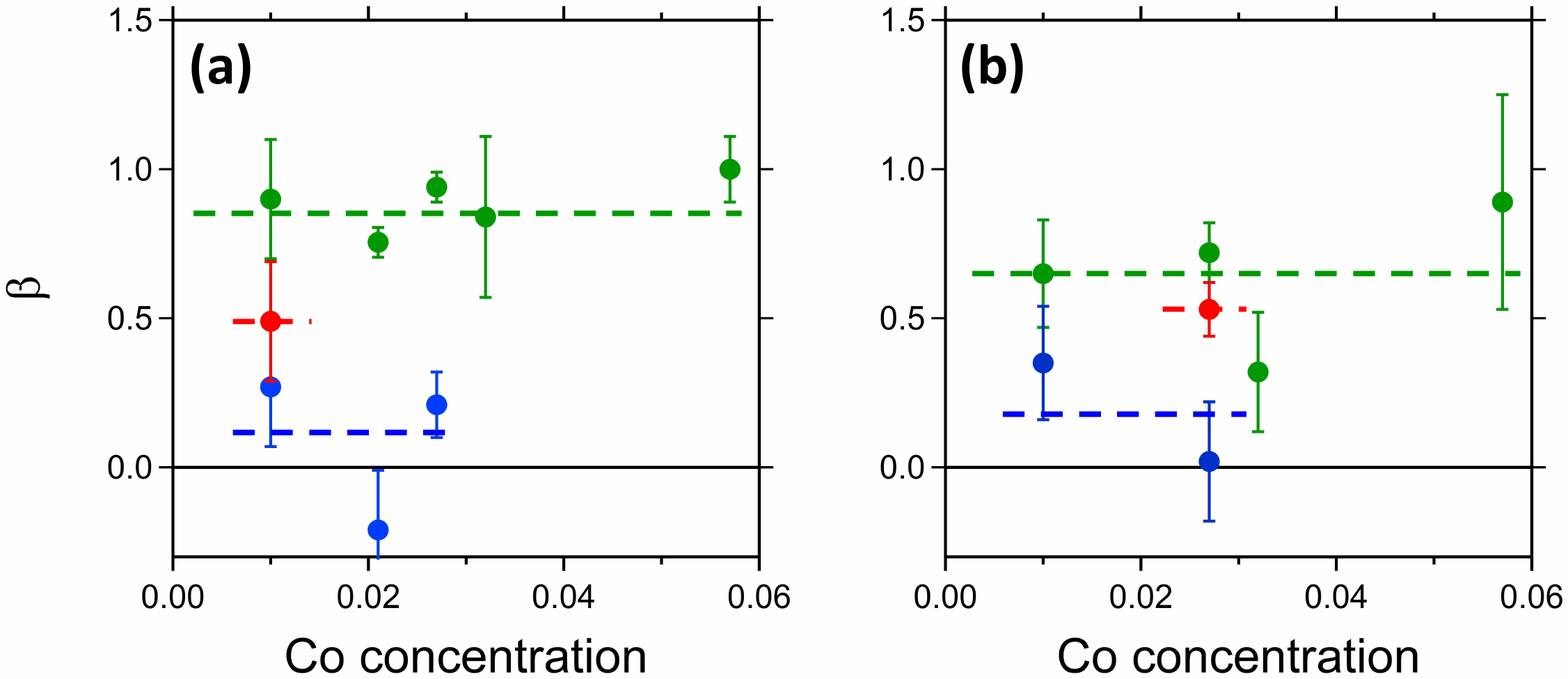}
\vspace{-2cm}
  \caption{
  (Color online) (a) Compilation of $\beta$ values of the hole pockets  of \NFCA\ as a function of Co concentration. (a) At three high symmetry points along cut I  (see Fig.\,1 of the main paper) near $Z=(0,0,\frac{\pi}{c})$. (b)At three symmetry points along cut II  along the diagonal of the Brillouin zone (see Fig.\,1 of the main paper). The color code corresponds to that used for the high-symmetry points in Fig.\,1 in the main paper.} 
 \label{mass}
 \centering
\end{figure}

In Fig.\,5(a) we present a compilation of the slopes of the scattering rates $\beta$ of  the hole pockets near $Z=(0,0,\frac{\pi}{c}$) of \NFCA\ as a function of the Co concentration. As in the case of $k_z$=0 (see Fig. \,3(d) of the main paper) within error bars we see no variation in $x$.
In Fig.\,5(b) we show an analogous compilation of $\beta$ related to a cut II along the diagonal of the Brillouin zone for the hole pockets of  \NFCA\ as a function of the Co concentration. Again, within error bars no dependence of $x$  is realized.
The fact that the average value of $\beta$ at point 1 is larger than that at point 3 possibly indicates an anisotropic scattering rate for the inner hole pocket. This anisotropy was also discussed in terms of anisotropic nesting conditions and anisotropic orbital character\,\cite{Kemper2011} as depicted in Fig.\,1 in the main paper. 

In Table I we compile the  $\beta$  values at the high-symmetry points at $k_z=0$ for the investigated four ferropnictide systems near optimal "doping" together with values for $\lambda_{MF}$ calculated from the $\beta$ values using the  effective bare masses $m_0$ from DFT calculations\,\cite{Singh2008,Deng2009}. The coupling constants for the 111 systems are considerably higher than those for the 122 compounds. This result is at present unclear and more systematic studies are necessary to clarify this point. 
 
\begin{table}[!tbp]
 \caption{ Slopes of the scattering rates $\beta$ of the investigated ferropnictides \BFAP\ (BFAP), \EFAP\ (EFAP),      \NFCA\ (NFCA), and \NFRA\ (NFRA) near optimal "doping" together with calculated coupling constants  $\lambda_{MF}$ .}
	\centering
		\begin{tabular}{c c c c c c c c c }
			\hline\hline
			Point & \multicolumn{2}{l}{ \hspace{1mm} BFAP}&\multicolumn{2}{l}{\hspace{1mm} EFAP } &\multicolumn{2}{l}{ \hspace{1mm}NFCA  }&\multicolumn{2}{l}{\hspace{1mm} NFRA }\\
			\hline
			    &$\beta $& $\lambda_{MF}$&$\beta $&$\lambda_{MF}$&$\beta $&$\lambda_{MF}$&$\beta $&$\lambda_{MF}$\\
			\hline 
			1 & 0.75 & 1.5 &0.81&1.6 &1.0&6.3&0.58&2.7 \\
			2 & 0.22 & 0.96&  $\approx 0$&$\approx 0$& & &0.36&2.3  \\
			7 & &  &  & &0.69&5.1&0.46&3.4 \\
			3 & 0.92&1.83  & & &0.80 &8.1& &  \\
			4& 0.22&0.73 & & &0.11 & 0.70& &  \\
			8 & 0.14 &0.55 & & &0.51 &3.7 & &   \\
			5 & 0.68&0.44 & 0.57&0.84 &0.61 & & & \\
			6 & 0.87&  & & & & & &  \\
	\noalign{\smallskip}\hline\hline
		\end{tabular}
\end{table}

\begin{figure}[!tbp]
  \centering
%\hspace{7cm}
  \includegraphics[angle=-90,width=\columnwidth]{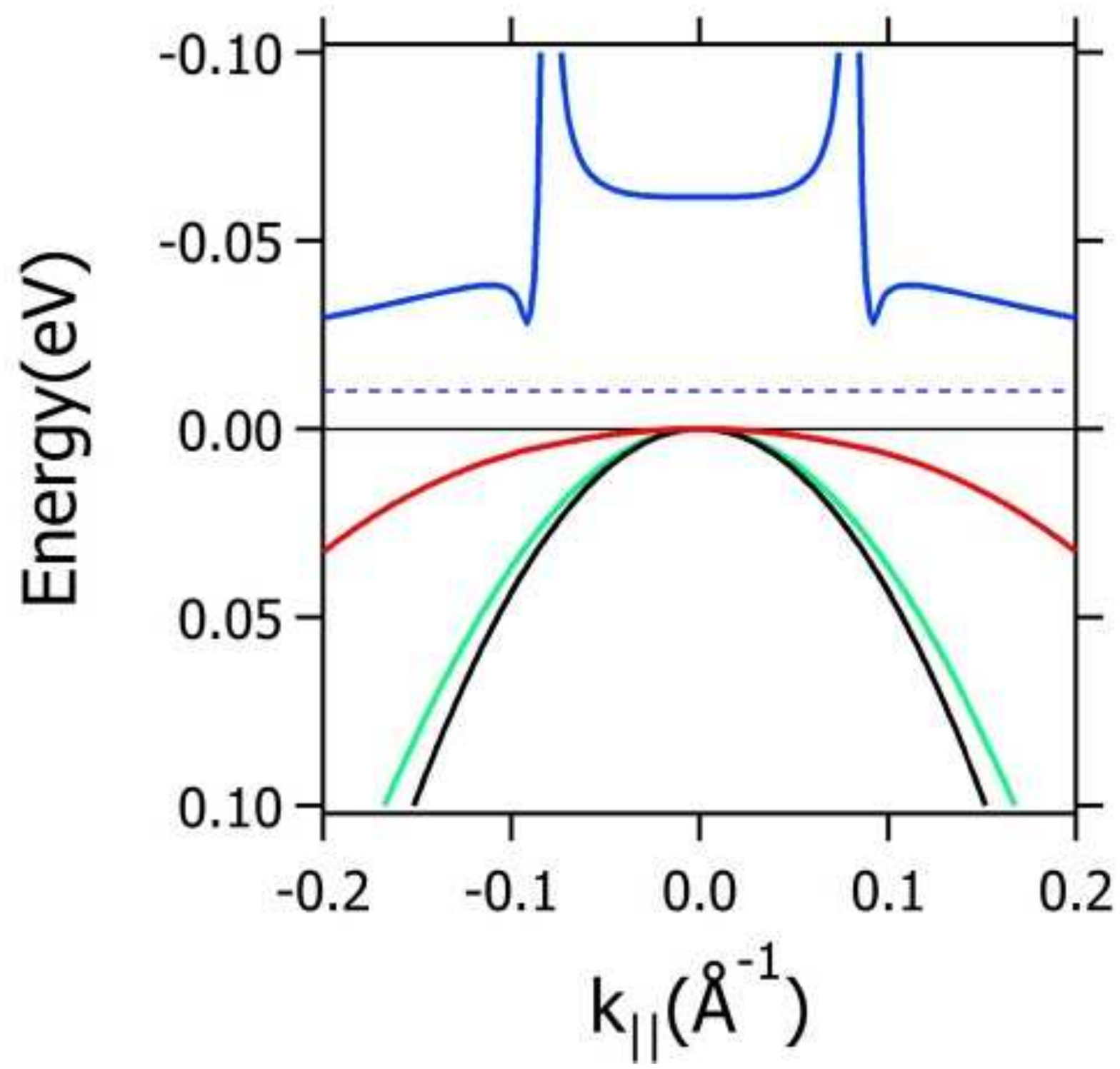}
  \caption{
  (Color online) Calculation of the renormalized dispersion (red line) using the bare particle dispersion from DFT calculations (black line) and the real part of the self-energy (green line) described in the text. In addition we show the effective mass $\frac{m^*}{m_0}$ (blue line) and the $\frac{m^*}{m_0}=1$  line (thin dashed blue line).} 
 \label{mass}
 \centering
\end{figure}

 From this compilation one realizes large differences of the scattering rates between the three hole pockets in \BFAP\  and a reduction of the differences when going to \NFCA\ and finally to
\NFRA\ . This evolution can be explained by the strength of the scattering potential of the substituents, being weak for the isovalent replacement of As by P, which is outside of the Fe layers, being stronger when Fe is replaced by Co, which occurs inside the Fe layer, and being strongest for replacement of Fe by the heavy element Rh.

\paragraph{Calculations for a Fermi-liquid/non-Fermi-liquid model.}
Here we present calculations in which we extend below $\omega_0$= 5 meV (the lower energy border in our ARPES results) the linear ARPES  $\Im\Sigma$  by a quadratic energy dependence (Fermi liquid behavior) . The result for the case where the top of the bare particle band touches the Fermi level is presented in Fig.\,6. The result is rather independent on the strength of the coupling constant of the Fermi liquid part. Comparing the present calculation with that presented in Fig.\,4(c) of the main paper there is still a weakly dispersing renormalized band at the Fermi level but  the effective mass at the $\Gamma$ point is now  reduced to $m^*/m_0\approx7$. The  strong mass enhancement now appears at $\omega_0$ since at this energy we have  a discontinuity in $\Im\Sigma$. Analogous calculations for the hole pocket shifted by 50 meV above the Fermi level (not shown) yield  that when the top of the bare particle hole pocket is not at the Fermi level, the renormalized dispersion is very close to that presented in Fig.\,4(d) of the main paper and the mass renormalization at the chemical potential  is reduced to $\approx 4$

These calculations  show that in both cases, the linear or the quadratic extension at low energies of the measured ARPES $\Im\Sigma $,
the crossing of  the top of the bare particle hole pocket plays an important role for the electronic structure of these correlated systems.  It can  explain the appearance of  control parameter  dependent (band shift dependent)  mass enhancement. Possibly the divergent mass enhancement in the transport properties observed in \BFAP\ near optimal substitution\,\cite{Analytis2014} or the strange behavior of  the thermal properties of \BFCA\ \,\cite{Meingast2012} can be explained in this model.

\bibliographystyle{phaip}
\bibliography{Pnictide}